# DFT based insights into elastic, thermophysical, electronic and optical properties of topological insulators XTe$_5$ (X = Zr, Hf)

Syed Shovon Mahbub Mahin[1], Suptajoy Barua[1], B. Rahman Rano[1*], Ishtiaque M. Syed[1], S. H. Naqib[2*]

[1]Department of Physics, University of Dhaka, Dhaka-1000, Bangladesh

[2]Department of Physics, University of Rajshahi, Rajshahi-6205, Bangladesh

*Corresponding author e-mails: rano167@du.ac.bd & salehnaqib@yahoo.com

**Abstract**

Transition metal penta-tellurides, ZrTe$_5$ and HfTe$_5$ have been recently drawn a lot of attention due to their fascinating physical properties and for being prominent materials showing topological phase transitions. In this study, we investigated mechanical, thermophysical and optoelectronic properties of these materials which remained almost unexplored till now. We also studied electronic properties and compared those with previous studies. We used Density Functional Theory (DFT) based calculations to study all of these properties. This study suggests that the materials are mechanically stable, possess high mechanical and bonding anisotropy and are brittle in nature. Our study also suggests that the compounds are soft in nature and they contain a mixture of covalent and metallic bonding. Investigation of thermophysical properties, namely, Grüneisen parameter and Debye temperature indicates weak bonding strength in these compounds. Analysis of melting temperature, thermal expansion coefficient, heat capacity, radiation factor, acoustic impedance, and minimum thermal conductivity suggests their possible application in acoustic and thermoelectric devices. Examination of their optical characteristics reveals that they have a considerable reflectivity from the infrared to the ultraviolet region. The refractive indices of these materials are high at low energy so they are potential candidates for reflective coating of solar radiation. There have been debates over exact topological natures of these compounds, whether they are semi-metals or insulators. Our study of electronic band structure and density of states reveal that spin-orbit interaction is responsible for enhancing energy gaps and promoting insulating characteristics in these compounds.

# 1. Introduction

Topological materials, topological phases of matter, their properties and possible applications have been dynamic and promising field of research in recent years. These materials can be classified into many different categories- topological insulators, topological Dirac semimetals, Weyl semimetals etc. [1,2].

Topological insulators (TIs) are materials with insulating bulk, while having time reversal symmetry protected conducting edge or surface states [2,3]. The discovery of both two-dimensional (2D) and three-dimensional (3D) TIs has attracted much attention in recent years due to underlying physics and their promising applications in electronic, spintronic and quantum computing devices [3–5]. TIs have shown excellent electronic and optical properties and they have been considered as feasible candidates of advanced optoelectronic devices, for example: plasmonic solar cell, nanometric holograms, near-infrared photodetectors etc [6–8]. Vigorous research on using TIs as saturable absorber in fiber laser technology, synthesizing TI thin films and to use them in designing topological superconductors have been going underway [9–13].

Transition metal penta-tellurides, $ZrTe_5$ and $HfTe_5$ have received considerable attention since their discovery [14,15]. They have been investigated since the 1970's because of their fascinating thermoelectric properties, being well known candidates to be used in thermoelectric devices, due to their anomalies in electrical transport phenomena, Hall coefficients, thermoelectric power, magnetic susceptibility and heat capacity [14,16–19]. Even depending on synthesis methods, these materials show distinct transport and thermoelectric properties [20]. In $ZrTe_5$, sign reversal in thermopower and resistivity peak at $T_p = 130$ K, has been known since 1980's [21]. In $HfTe_5$, this peak is found around 50-90 K [16,21,22]. Pressure driven superconductivity and structural phase transition were reported in these compounds and pressure evolution of critical temperature have been studied [23,24]. Further investigations on these compounds reveal more fascinating behaviors of these compounds, such as the chiral magnetic effect, log periodic quantum oscillations, discrete scale invariance, temperature induced Lifshitz transition, Kohn anomaly, chiral anomaly, magneto chiral anisotropy, negative magnetoresistance, 3D quantum Hall effect, unconventional Hall effect [25– 32]. The underlying mechanism of this mysterious resistivity anomaly, sign reversal of Hall, Seebeck co-efficient and thermoelectric power near transition temperature, have been attributed to structural phase transition, bipolar conduction, formation of charge/spin density wave, Lifshitz transition, polaronic behavior etc. [20,21,33–39]. Recently, interests in these materials have been boosted due to their exotic topological properties and temperature dependent topological phase transition [34,40]. In 2014, a study predicted the mono-layer of $ZrTe_5$ and $HfTe_5$ to be quantum spin Hall insulators [41]. The bulk counterparts were predicted to be 3D TIs located at the phase transition boundary between the strong and weak TI [41,42]. Until now, different experimental works concluded different findings on these compounds. In these works, they are predicted to be either Dirac semi-metal, strong TI or weak TI. [42]. Some magneto-infrared spectroscopy and angle-resolved photoemission spectroscopy (ARPES) studies show that $ZrTe_5$ is a Dirac semimetal [43–45]. On the other hand, $ZrTe_5$ and $HfTe_5$ are predicted to be small gap TIs by some recent STM and ARPES results and electrical transport measurements [46–50]. Presence of spin polarized states near Fermi level and subsequent ARPES study suggests strong TI phase of $ZrTe_5$ whereas several high-resolution laser ARPES measurements, Shubnikov-De-Hass oscillation support the weak TI nature [46,48,51–53]. Recent STM and STS studies also suggest weak TI nature of these compounds [54]. $ZrTe_5$ even has been reported as a nodal line semimetal and magnetic field induced flat bands have been observed [55]. Experimental manipulation to host a multitude of 1D topological edge channels through anisotropic surface of ZrTe$_5$ have been proposed, suggesting this material to be deployed in low-power-consumption electronic nano device [56]. Over the past decades, the anomalies in transport properties, sign reversal of Hall and Seebeck coefficient, splitting of Landau levels due to Zeeman effect have been extensively studied, but their relation with band topology has drawn attention very recently [57,58]. All of these interesting results make these compounds potential candidates for studying novel physical phenomena in detail.

To the best of our knowledge, elastic and optical properties of these materials are still unexplored. Studies of several thermophysical properties of these compounds are also lacking. Elastic and thermophysical properties can help us to understand the possibility of applying these compounds in industrial sectors. A comprehensive understanding of electronic and optical properties is essential to investigate their possible application in electronic and optoelectronic devices. These motivated us to conduct our studies. Our aim is to study several bulk properties such as elastic and optical properties of these materials. At first, we optimized the geometry of the crystals to achieve ground state geometry of these compounds. Then we calculated single crystal elastic constants and compliance constants, by which we successively calculated several structural and thermophysical bulk properties such as, elastic moduli, anisotropic factors, Debye temperature, sound velocity, thermal expansion co-efficient, thermal conductivity etc. We also calculated the band structure and density of states with and without spin orbit interaction (SOI). Electronic properties with SOI suggested insulating nature of these materials. Then we calculated several optical properties which reveals various possibility of using them in optoelectronic applications.
The rest of this paper has been structured as follows: the computational methodology has been discussed in brief in section 2, the results and analyses are presented in section 3. Finally, the key findings of our study have been discussed in section 4.

## 2. Computational Methodology

We have used Density Functional Theory (DFT) based calculations, also known as *ab-initio* approach, to study the physical properties of the systems. We solve for an interacting many body system using the Kohn-Sham scheme [59,60]. In our study, we used CAmbridge Serial Total Energy Package (CASTEP) and Vienna Ab initio Simulation Package (VASP), two well-known DFT based simulation packages [61,62]. At first, we performed geometry optimization using CASTEP to find ground state geometry of the systems. To find more reliable approximation of exchange correlation effects, we used both Local Density Approximation (LDA) and Generalized Gradient Approximation (GGA) of Perdew-Burke-Ernzerhof (PBE) scheme [63,64]. For both of the compounds, GGA provided better result compared to the experimental data available. While using CASTEP, to model the electron-ion interactions, norm-conserving pseudopotentials have been used, with the Koelling-Harmon relativistic treatment. This type of pseudopotentials replicate the scattering properties of actual potential pretty successfully [65,66] . Broyden Fletcher Goldfarb Shanno (BFGS) algorithm has been used in geometry optimization [67]. For the self-consistent calculations, density mixing electronic minimizer has been used. The following convergence tolerances, were set in geometry optimization: $5\times10^{-6}$ eV/atom for energy, maximum 0.01 eV/Å force, highest 0.0005 Å displacement and maximum 0.02 GPa stress. Cut-off energy for plane wave basis set for ZrTe$_5$ and HfTe$_5$ were taken to be 550 eV and 450 eV, respectively. This ensured satisfactory level of total energy convergence. To sample the special k-points of the Brillouin Zone (BZ), a mesh size of 14×4×4 (28 irreducible k points) within the Monkhorst-Pack scheme has been set for both of these compounds [68]. For electronic and optical properties calculations, a mesh size of 18×5×5, i.e., separation 0.003 1/Å, were used for both systems. In case of pseudo atomic calculations of ZrTe$_5$ and HfTe$_5$, [$4s^2$ $4p^6$ $4d^2$ $5s^2$] for Zr, [$5d^2$ $6s^2$] for Hf and [$5s^2$ $5p^4$] for Te were treated as valence electron orbitals. Variety of studies on different materials including topological systems, have suggested that bulk elastic constants, thermophysical properties, several elastic modulus and bulk optical properties are affected insignificantly by spin-orbit interaction (SOI) [69–74]. So, SOI is not considered in calculating these properties. However, band structure and density of states (DOS) are affected by SOI.

For both of the systems, we also used VASP to relax the structure at first and prepare it for further calculations. Then we calculated electronic and optical properties. Using VASP, we were able to calculate the band structures of these compounds incorporating SOI. For standard relaxation we used energy cut-off 550 eV and 450 eV respectively for ZrTe$_5$ and HfTe$_5$. Precision level was set to accurate, energy difference for self-consistent field (SCF) energy convergence was set to be $10^{-8}$ eV and threshold force which corresponds to condition for truncating ionic relaxation loop was set to be -0.01 eV/ Å. We didn't include smearing and the sigma value was set to 0.05 for all the calculations, but for DOS calculation, tetrahedron method with Blöchl corrections was used. For standard relaxation, a mesh size of 8×2×2 and for static calculations (i.e., band structure, DOS and optical properties), a mesh size of 25×6×7 was used. VASP uses projector-augmented wave (PAW) method to describe electron-core interaction [75,76]. We again used GGA-PBE functional and chose pseudopotential files Zr_sv, Hf_pv and Te so that [$4s^2$ $4p^6$ $4d^3$ $5s^1$] for Zr, [$5p^6$ $5d^3$ $6s^1$] for Hf and [$5s^2$ $5p^4$] for Te were treated as valence electron orbitals.

## 3. Result and Analysis

### 3.1. Structural Features

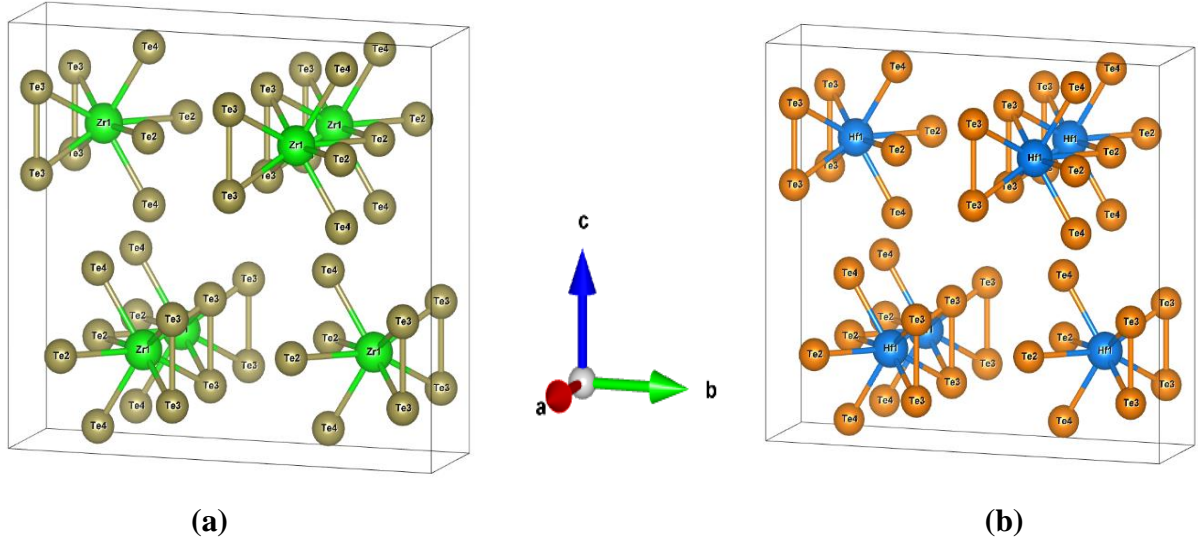

**(a)** **(b)**

**Figure 1:** Schematic crystal structures of (a) $ZrTe_5$ & (b) $HfTe_5$.

$XTe_5$ (X= Zr, Hf) crystalizes in orthorhombic structure in ground state and belongs to the space group Cmcm (no. 63). The X1 atom attains 4c position. The crystal contains triagonal prismatic rods as building block. These prisms are constructed by two types of Te atoms, Te2 and Te3, which stay at the Wyckoff positions of 4c and 8f, respectively. Other Te atoms, which are referred to here as Te4, form zigzag chains with the X1 atoms [15,40]. Initial atomic positions and lattice parameters for our simulation were taken from previous experimental and theoretical works [15,77].

In Table 1, we have presented the obtained lattice parameters and cell volume by GGA and LDA calculations. Previously reported values are also presented here and we compared our results with one of the experimental works. Results using GGA provide better agreement when compared to experimental values. So, for further calculations, we used GGA as our exchange-correlation energy functional.

**Table 1:** Comparison of calculated lattice parameters and cell volumes of $XTe_5$ with previous works. The lattice parameters a, b, and c are in Å, the unit cell volume (V) is in Å$^3$ and the error (in %) of volume ($\%\delta V/V$) is compared with reference work [15].

| Material | a | b | c | V | % δV/V | Functional & Reference Work |
|---|---|---|---|---|---|---|
| $ZrTe_5$ | 4.0116 | 14.8126 | 13.4113 | 796.9550 | 1.19 | GGA (This work) |
| | 3.9238 | 14.0985 | 13.3927 | 740.8848 | 5.93 | LDA (This work) |
| | 3.9797 | 14.4700 | 13.6760 | 787.5496 | - | Experimental Work [15] |
| | 3.9830 | 14.4933 | 13.7003 | 790.8300 | - | Experimental Work [20] |
| | 3.9813 | 14.5053 | 13.7030 | 791.3512 | - | Experimental Work [20] |
| | 3.9800 | 14.5000 | 13.7300 | - | - | Experimental Work [19] |
| $HfTe_5$ | 3.9517 | 14.6350 | 13.3254 | 770.6650 | 1.63 | GGA (This work) |
| | 3.8917 | 14.0985 | 13.3714 | 731.3218 | 6.65 | LDA (This work) |
| | 3.9640 | 14.4430 | 13.6840 | 783.4370 | - | Experimental Work [15] |
| | 3.9741 | 14.4812 | 13.7202 | - | - | Experimental Work [24] |
| | 4.0895 | 15.7709 | 13.8831 | - | - | GGA (Theoretical Work) [78] |
| | 3.9227 | 14.3026 | 13.5120 | - | - | LDA (Theoretical Work) [78] |

## 3.2. Elastic Properties

The mechanical and dynamical properties of materials are heavily linked with elastic constants. We used stress-strain method implemented by CASTEP to calculate single crystal elastic constants. According to *Voigt notation*, the elastic constants of a crystal form a 6 × 6 symmetric matrix. This matrix is called stiffness matrix [79]. Number of symmetry operations of crystal system determines the number of independent components of this matrix. Orthorhombic structure has nine independent components because of symmetry constraints. We have tabulated our obtained results of single-crystal elastic constants along with elastic compliance constants for both compounds in Table 2.

**Table 2:** Calculated single-crystal elastic constants ($C_{ij}$ in GPa) and elastic compliance constants ($S_{ij}$ in 1/GPa) for $ZrTe_5$ and $HfTe_5$.

| Components | ZrTe$_5$ | | HfTe$_5$ | |
|---|---|---|---|---|
| *ij* | C | S | C | S |
| 11 | 78.9045 | 0.0136343 | 83.2424 | 0.0130265 |
| 22 | 36.6130 | 0.0279056 | 37.6673 | 0.0269591 |
| 33 | 75.8126 | 0.0142346 | 82.3512 | 0.0131914 |
| 44 | 5.3768 | 0.1859825 | 4.0402 | 0.2475094 |
| 55 | 34.7669 | 0.0287630 | 38.4851 | 0.0259840 |
| 66 | 9.6802 | 0.1033031 | 8.6207 | 0.1159992 |
| 12 | 5.8267 | -0.0015748 | 5.2453 | -0.0012882 |
| 13 | 19.6224 | -0.0033957 | 22.4342 | -0.0034591 |
| 23 | 6.4151 | -0.0019537 | 5.7254 | -0.0015234 |

Necessary and sufficient conditions for mechanical stability of a crystal system are known as *Born stability criteria* [80,81]. For orthorhombic system in the ground state, the Born conditions for mechanical stability are [81]:

$$C_{11} > 0;\ C_{11}C_{12} > C_{12}^2;$$
$$C_{11}+C_{22}+C_{33} + 2C_{12}C_{13}C_{23} - C_{11}C_{23}^2 - C_{22}C_{13}^2 - C_{33}C_{12}^2 > 0; \qquad (1)$$
$$C_{44} > 0;\ C_{55} > 0;\ C_{66} > 0$$

For both systems, all the inequalities are satisfied, indicating their mechanical stability.

$C_{11}$, $C_{22}$ and $C_{33}$, the diagonal entries of the stiffness matrix are measurements of material's ability to resist tensile stress applied respectively in the crystallographic directions **a**, **b** and **c**. For our compounds, $C_{22}$ are less than $C_{11}$ and $C_{33}$, which indicates more compressibility in crystallographic **b** direction. Within the **a-c** plane, bonding strength is higher. The response of the crystal to shear are characterized by $C_{44}$, $C_{55}$, and $C_{66}$. For example, $C_{44}$ measures the resistance against shear across the (100) plane in the [010] direction, when applied stress is tangential. Low value of $C_{44}$ indicates the materials' inability to withstand shear deformation in (100) plane. The small value of the last three diagonal constants compared to the first three signifies that the crystals are less resistant to shearing strain.

Rest of the constants $C_{12}$, $C_{13}$ and $C_{23}$, represent the ability of a crystal to resist orthogonal distortions with the constraint of volume conservation. $C_{13}$ is basically a combination of a functional stress component along the crystallographic **a** direction and uniaxial strain along the crystallographic **c** direction.

We have calculated Cauchy pressure ($C''$) for our systems. For an orthorhombic crystal, the Cauchy pressure can be defined as $C''_1 = (C_{23} - C_{44})$ for (100) plane, $C''_2 = (C_{13} - C_{55})$ for (010) plane, and $C''_3 = (C_{12} - C_{66})$ for (001) plane [82]. Cauchy pressure can be helpful to investigate the nature of atomic bonding in a material and its ductility or brittleness. Positive value of Cauchy pressure corresponds to the dominance of non-directional metallic bonding whereas negative value corresponds to the dominance of directional covalent nature of bonding. Also, positive value of Cauchy pressure suggests ductile nature of the compound, while negative value suggests brittle nature [86,87].

Another useful structural parameter is the Kleinman parameter, $\zeta$, which describes a compound's stability under bending and stretching and generally resides between 0 and 1. Values close to 0 signifies relative ease in bond stretching and values close to 1 suggest ease in bond bending [85]. We have calculated Kleinman parameter using the following formula [86]:

$$\zeta = \frac{C_{11} + 8C_{12}}{7C_{11} + 2C_{12}} \qquad (2)$$

The tetragonal shear modulus, calculated via the equation given by,

$$C' = \frac{C_{11} - C_{12}}{2} \qquad (3)$$

is a useful parameter to measure the stiffness of a crystal. It is also an important parameter related to long wavelength transverse acoustic waves and important for structural transformation of a crystal [87].

In Table 3, we have presented parameters for both of the compounds: tetragonal shear modulus, Kleinman parameter and Cauchy pressures.

**Table 3:** Cauchy pressures ($C''$ in GPa), Kleinman parameter ($\zeta$) and Tetragonal shear modulus ($C'$ in GPa) for $ZrTe_5$ and $HfTe_5$.

| Material | $C''_1$ | $C''_2$ | $C''_3$ | $\zeta$ | $C'$ |
|---|---|---|---|---|---|
| $ZrTe_5$ | 1.0383 | -15.1445 | -3.8535 | 0.2225 | 36.5389 |
| $HfTe_5$ | 1.6852 | -16.0509 | -3.3754 | 0.2111 | 38.9986 |

Bonding anisotropy is present in our compounds as confirmed from different values of Cauchy pressure for different planes. For (100) plane, the chemical bonding is metallic, but for other two planes, there is a clear dominance of covalent bonding. The values suggest overall brittle nature of the compounds due to dominance of directional bonding. The low values of the Kleinman parameter suggest that bond stretching contribution is greater than bond bending contribution in mechanical strength of these compounds. Tetragonal shear modulus values suggest these compounds are comparatively stiffer [73,88].

We can use our obtained single crystal elastic constants to calculate several polycrystalline elastic moduli and some other bulk parameters. We have calculated bulk modulus ($B$), shear Modulus ($G$), Pugh's ratio ($B/G$), Young's modulus ($Y$), Poisson's ratio ($v$), machinability index ($\mu_M$), Vicker's hardness ($H_v$), microhardness ($H_{micro}$) and macrohardness ($H_{macro}$). All of these results are tabulated in Table 5 and Table 6. For elastic moduli calculations, Voigt approximation assumes strain distribution is continuous throughout the material. It also assumes discontinuity in stress distribution and provides an upper bound on elastic moduli [89]. On the other hand, Reuss approximation, assumes continuous stress with discontinuity in strain distribution inside the grains, providing lower bound on elastic moduli [90]. However, the closest approximation to the true polycrystalline elastic moduli, Hill's approximation is found by the arithmetic average of the two values obtained from the respective Voigt and Reuss approximations [91]. Bulk modulus measures the ability of a compound to resist volume change when isotropic pressure is applied. Shear modulus is the ratio of tangential stress and shear strain. It quantifies the resistance to plastic deformation. They are also useful measurements of bonding strength of a material. The formulas we have used are given below [91]:

$$B_R = \frac{1}{(S_{11}+S_{22}+S_{33})+2(S_{12}+S_{13}+S_{23})} \quad (4)$$

$$B_V = \frac{1}{9}(C_{11}+C_{22}+C_{33}) + \frac{2}{9}(C_{12}+C_{13}+C_{23}) \quad (5)$$

$$G_R = \frac{15}{4(S_{11}+S_{22}+S_{33})-4(S_{12}+S_{13}+S_{23})+3(S_{44}+S_{55}+S_{66})} \quad (6)$$

$$G_V = \frac{1}{15}(C_{11}+C_{22}+C_{33}-C_{12}-C_{13}-C_{23}) + \frac{1}{5}(C_{44}+C_{55}+C_{66}) \quad (7)$$

$$B_H = \frac{B_R+B_V}{2} \quad \text{and} \quad G_H = \frac{G_R+G_V}{2} \quad (8)$$

The ratio between these two moduli, $B/G$ is known as Pugh's ratio and can be used to determine if a material is ductile or brittle [92]. The material is ductile if the ratio is greater than the critical value (1.75), otherwise it is brittle in nature [92,93]. Young's modulus, $Y$ is defined as the ratio of the tensile stress to the longitudinal strain. It is a measure of a material's stiffness, describing its ability to resist deformation under stress. It is crucial in engineering and materials science for designing structures and predicting elastic behavior. It is calculated using this formula [94]:

$$Y = \frac{9BG}{3B+G} \tag{9}$$

Poisson's ratio (*v*) is a measure of materials deformation along the perpendicular direction of loading. It is a useful parameter which describes compound's ductility/brittleness, stability against shear and characteristics of bonding force of a material. If $v$ is greater (less) than 0.26, the material is ductile (brittle) [95,96]. Poisson's ratio resides within the range, $-1.0 \leq \nu \leq 0.50$. but, specially, in between 0.25 and 0.50 for solids where central forces are dominating [94]. This ratio is also useful in predicting the nature of bonding in a compound. Values close to 0.10 suggest pure covalent nature of chemical bonding, whereas, values above 0.33 suggest metallic bonding [97]. We have used the following formula to calculate Poisson's ratio [93,98]:

$$\nu = \frac{3B-2G}{2(3B+G)} \tag{10}$$

The ratio of bulk modulus and $C_{44}$ is known as machinability index ($\mu_M$) which is an important parameter in engineering application of a material [93,99]. It measures the ease with which a material can be machined using cutting/shaping tools. A high value of the machinability index implies better dry lubricating properties and low machining cost. We have also calculated several other hardness parameters. Microhardness is useful in characterization of solids. This parameter is related to the ionicity, the bond-length, the valence electron density etc. Dislocations and other point defects have a noticeable impact on microhardness [100–102]. Macrohardness is another useful parameter when we use loads which are comparatively bigger and heavier. We also calculated Vicker's hardness which is a useful measurement of hardness for both soft and hard materials [103,104]. Materials with a high value of hardness parameters are comparatively difficult to machine. But they are comparatively more useful for cutting tools and wear-resisting coatings. The formulas we have used are as follows [103,105,106]:

$$\begin{aligned} H_V &= 0.92(G/B)^{1.137}G^{0.708} \\ H_{micro} &= \frac{(1-2v)E}{6(1+v)} \\ H_{macro} &= 2[G(G/B)^2]^{0.585} - 3 \end{aligned} \tag{11}$$

**Table 4:** Using Voigt, Reuss, and Hill's approximations, the isotropic bulk modulus (*B* in GPa), shear modulus (*G* in GPa), Pugh's ratio ($B_H/G_H$) and for ZrTe$_5$ and HfTe$_5$ obtained from the single crystal elastic constants.

| Material | $B_R$ | $B_V$ | $B_H$ | $G_R$ | $G_V$ | $G_H$ | $B_H/G_H$ |
|---|---|---|---|---|---|---|---|
| ZrTe$_5$ | 23.851 | 28.339 | 26.095 | 12.448 | 20.595 | 16.552 | 1.579 |
| HfTe$_5$ | 24.608 | 30.007 | 27.308 | 10.666 | 21.552 | 16.109 | 1.695 |

**Table 5:** Young's modulus ($Y$ in GPa), Poisson's ratio ($v$), Machinability index ($\mu_M$) and several Hardness parameters ($H_V$, $H_{micro}$, $H_{macro}$ in GPa), from Hill's approximation, for ZrTe$_5$ and HfTe$_5$.

| Material | $Y$ | $v$ | $\mu_M$ | $H_V$ | $H_{micro}$ | $H_{macro}$ |
|---|---|---|---|---|---|---|
| ZrTe$_5$ | 40.989 | 0.238 | 4.853 | 3.991 | 2.887 | 3.052 |
| HfTe$_5$ | 40.386 | 0.254 | 6.759 | 3.612 | 2.653 | 2.483 |

For both of the compounds, we got $B_H > G_H$, which indicates that the mechanical failure of these compounds is controlled by applied shear components. Lower values of elastic moduli compared to other materials [71,73,74,107–109] suggest that our systems are comparatively soft in nature. Pugh's ratio suggests brittle nature, consistent with Cauchy pressure measurement. Lower values of Young's moduli indicate their inability to resist large tensile stress. Poisson's ratio again confirms brittle nature of both compounds, also suggests the domination of non-central force in ZrTe$_5$, and central force in HfTe$_5$. The values also suggest that there is a mixture of metallic and covalent bonding in our systems, with unequal proportion of metallic and covalent contribution. Low values of $v$ also indicate large volume change associated with elastic deformation.

These materials with a combination of metallic and covalent bondings, show very high level of machinability [73,110]. This also indicates excellent lubricating properties, low machining cost and lower friction. So, the compounds are easily machinable and useful in device fabrication. Low values of hardness parameters also support this result. Also, HfTe$_5$ has lower hardness and larger machinability compared to ZrTe$_5$.

For completeness we calculated directional bulk moduli and elastic anisotropic factors. We have also calculated $\alpha$ and $\beta$ which are defined as the relative change of the **b** and **c** axis, respectively, as a function of the deformation of the **a** axis [93]. These help to understand anisotropies present in bonding and other mechanical properties of these compounds. For orthorhombic crystal systems, the bulk modulus can be calculated along the crystallographic axes using the single crystal elastic constants we calculated [93]. By considering the definition of the bulk modulus where the strains perpendicular to the stress directions are all equal, which is the response due to hydrostatic pressure, we get a formula for lower bound of bulk modulus ($B_{relax}$) [93]. We can also calculate upper bound of bulk modulus ($B_{unrelax}$) using single crystal elastic constants [93]. All of these are calculated and shown below in Table 6.

**Table 6:** $\alpha$, $\beta$, bulk modulus ($B_{relax}$ in GPa) and its upper bound ($B_{unrelax}$ in GPa), bulk modulus along the crystallographic axes $a$, $b$, $c$ ($B_a$, $B_b$, $B_c$ in GPa) for ZrTe$_5$ and HfTe$_5$.

| Material | $\alpha$ | $\beta$ | $B_{relax}$ | $B_{unrelax}$ | $B_a$ | $B_b$ | $B_c$ |
|---|---|---|---|---|---|---|---|
| ZrTe$_5$ | 2.8137 | 1.0256 | 23.8515 | 28.3398 | 115.4244 | 41.0223 | 112.5433 |
| HfTe$_5$ | 2.9167 | 0.9915 | 24.6089 | 30.0079 | 120.7856 | 41.4175 | 121.8211 |

The values of $B_{relax}$ and $B_{unrelax}$ are same as $B_R$ and $B_V$ calculated earlier. So, the calculations are consistent. Unequal values of bulk moduli along different crystallographic directions signify bonding anisotropy in these materials. $B_b$ is lower than other two directional moduli, this indicates the materials are highly compressible when stress is applied along **b** direction. The anisotropy is strongest along **b** direction with respect to the one within **a-c** plane. These results reconfirm our earlier findings.

Elastic anisotropy information is essential for materials engineering. Microcracks formation in a crystal and crystal's durability are heavily linked with elastic anisotropy in crystals [93, 111]. The shear anisotropy factors ($A_i$) provide a criterion to measure degree of anisotropy in bonding strength in different atomic planes. The shear anisotropy factor for the (100) shear planes between the [011] and [010] directions; for the (010) shear planes between [101] and [001] directions, and finally, for the (001) shear planes between [110] and [010] directions are given, respectively, by [95]:

$$A_1 = \frac{4C_{44}}{C_{11}+C_{33}-2C_{13}} \tag{12}$$

$$A_2 = \frac{4C_{55}}{C_{22}+C_{33}-2C_{23}} \tag{13}$$

$$A_3 = \frac{4C_{66}}{C_{11}+C_{22}-2C_{12}} \tag{14}$$

Measurements of anisotropy parameters of the bulk modulus along **a** axis and **c** axis with respect to the **b** axis are given by:

$$A_{B_a} = \frac{B_a}{B_b} = \alpha \quad \text{and} \quad A_{B_c} = \frac{B_a}{B_c} = \frac{\alpha}{\beta} \tag{15}$$

The anisotropy in compressibility and shear:

$$A_B = \frac{B_V - B_R}{B_V + B_R} \quad \text{and} \quad A_G = \frac{G_V - G_R}{G_V + G_R} \tag{16}$$

Universal anisotropy ($A^U$) is also calculated using the relation [112]:

$$A^U = 5G/G_R + B_V/B_R - 6 \tag{17}$$

**Table 7:** Universal anisotropy factor, $A^U$, anisotropy parameters related to elastic moduli: $A_G$, $A_B$, $A_{Ba}$, $A_{Bc}$, shear anisotropy factors: $A_1$, $A_2$, $A_3$ for ZrTe$_5$ and HfTe$_5$.

| Material | $A^U$ | $A_G$ | $A_B$ | $A_{Ba}$ | $A_{Bc}$ | $A_1$ | $A_2$ | $A_3$ |
|---|---|---|---|---|---|---|---|---|
| **ZrTe₅** | 3.4606 | 0.2465 | 0.0859 | 2.8137 | 2.7435 | 0.1863 | 1.3963 | 0.3728 |
| **HfTe₅** | 5.3225 | 0.3378 | 0.0988 | 2.9167 | 2.9417 | 0.1338 | 1.4179 | 0.3123 |

The values of shear anisotropic factors and universal anisotropy factor indicates the presence of elastic and mechanical anisotropy in a large extent. $A_G$ and $A_B$ lie between 0 and 1, where 0 stands for isotropy in bulk and shear moduli. The results for both compounds suggest that anisotropy in shear moduli (24.65% and 33.78%) is larger than anisotropy in bulk moduli (8.59% and 9.88%).

Directional dependence of different elastic moduli and Poisson's ratio in three dimensional systems can be visualized using ELATE [113, 114]. If the shapes are perfect sphere, the compounds are elastically isotropic. Deformed shapes other than sphere suggest anisotropy in elastic properties. For our systems, the results are given below from which anisotropy is clearly evident:

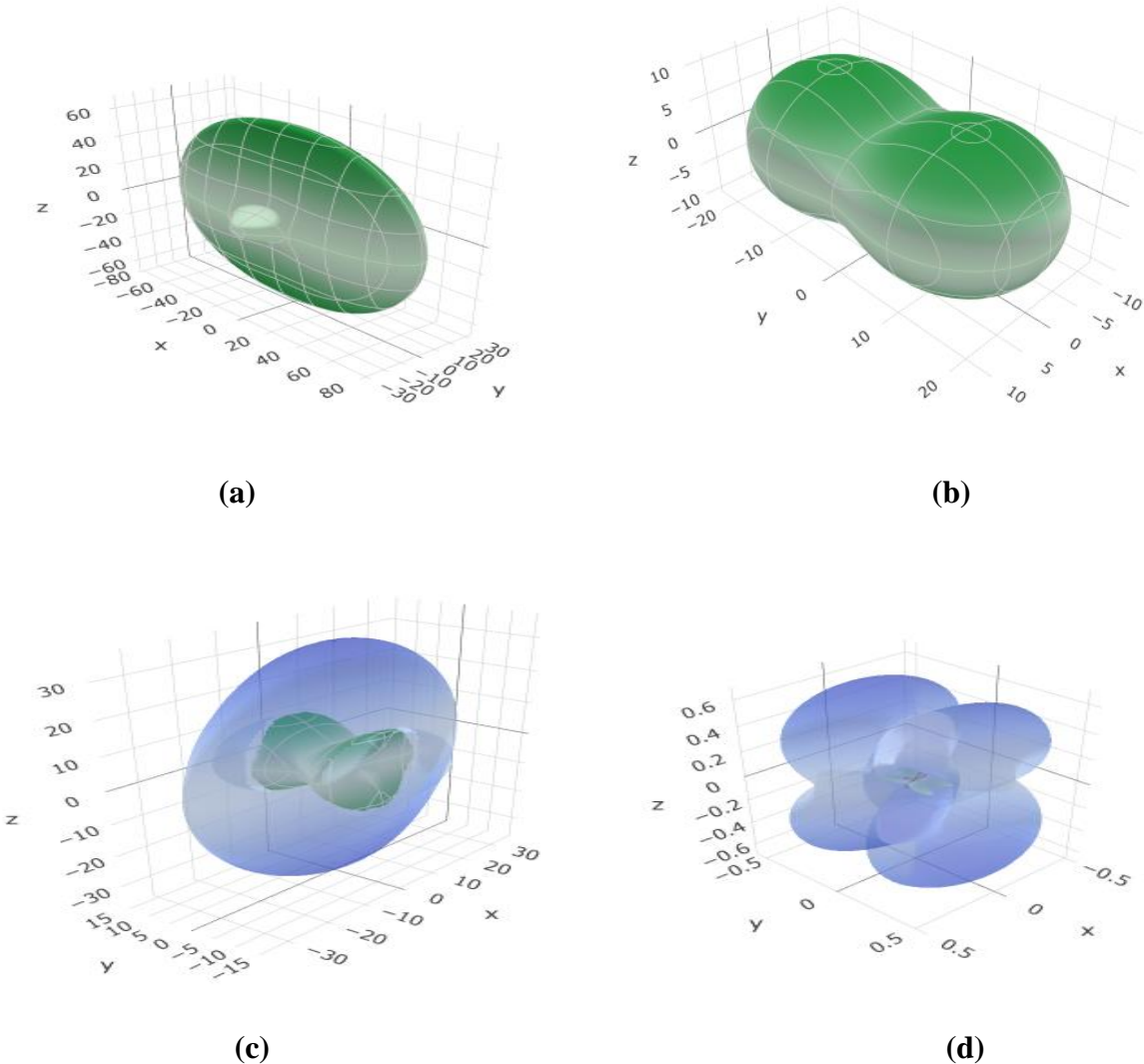

**Figure 2:** (a) Young's modulus (b) Linear compressibility (c) Shear modulus (d) Poisson's ratio in different directions for $ZrTe_5$

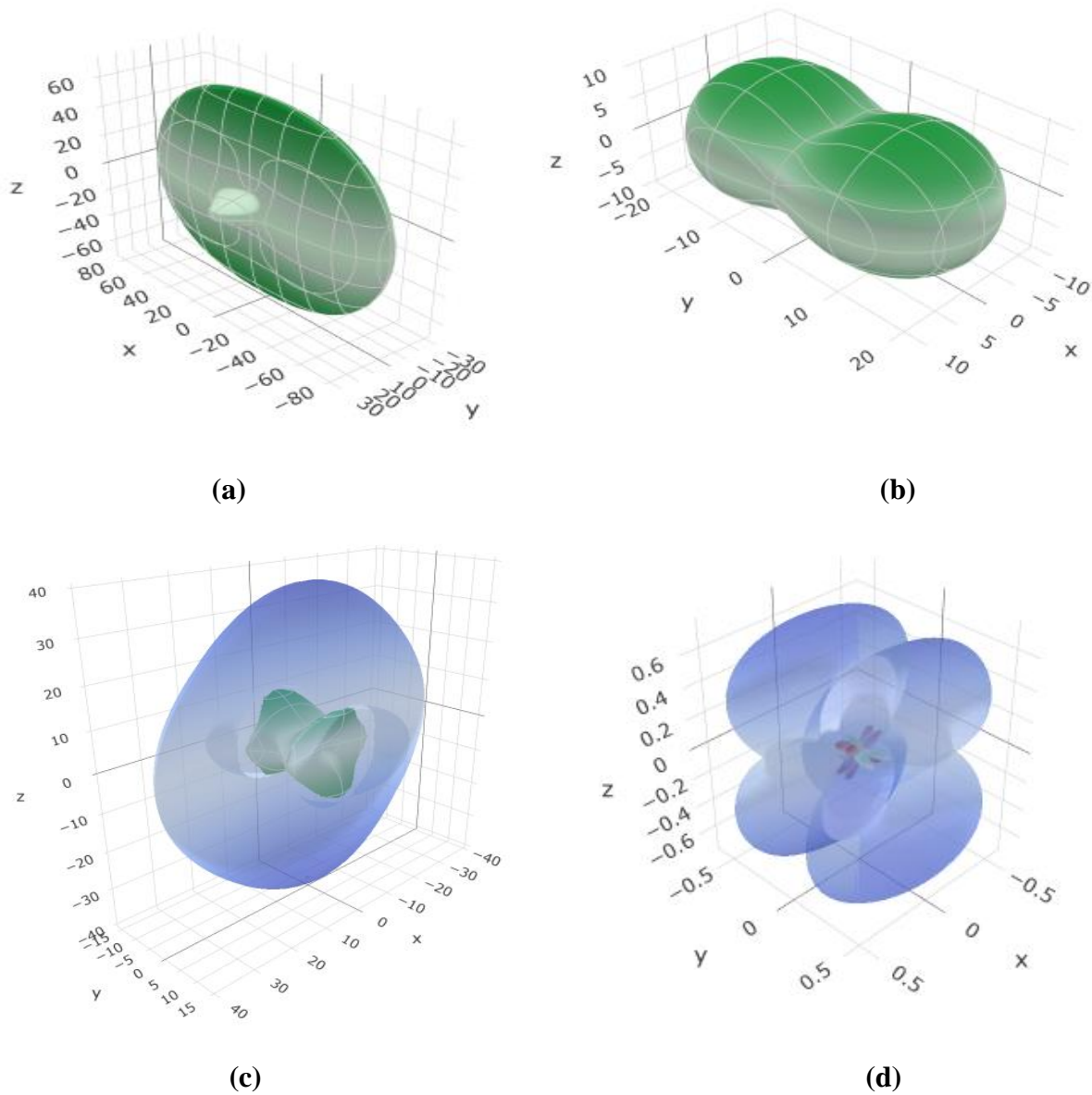

**Figure 3:** (a) Young's modulus (b) Linear compressibility (c) Shear modulus (d) Poisson's ratio in different directions for HfTe$_5$

## 3.3. Thermophysical Properties

Elastic constants are closely connected to elastic wave velocities through materials. At first, we calculated density of the compounds and then using calculated elastic constants, we calculated anisotropic sound velocities for different propagation directions. We have used the formulas [115]:

| Propagation Direction | Longitudinal velocity | Transverse Velocity-I | Transverse Velocity-II |
|---|---|---|---|
| [1 0 0] | $\sqrt{\frac{C_{11}}{\rho}}$ | $\sqrt{\frac{C_{66}}{\rho}}$ | $\sqrt{\frac{C_{55}}{\rho}}$ |
| [0 1 0] | $\sqrt{\frac{C_{22}}{\rho}}$ | $\sqrt{\frac{C_{66}}{\rho}}$ | $\sqrt{\frac{C_{44}}{\rho}}$ |
| [0 0 1] | $\sqrt{\frac{C_{33}}{\rho}}$ | $\sqrt{\frac{C_{55}}{\rho}}$ | $\sqrt{\frac{C_{44}}{\rho}}$ |

The obtained results are tabulated below in Table 8.

**Table 8:** The density ($\rho$ in g/cm$^3$) and sound velocities for different propagation directions ($v_l$, $v_{t1}$, $v_{t2}$ in m/s).

| Material | $\rho$ | [1 0 0] | | | [0 1 0] | | | [0 0 1] | | |
|---|---|---|---|---|---|---|---|---|---|---|
| | | $v_l$ | $v_{t1}$ | $v_{t2}$ | $v_l$ | $v_{t1}$ | $v_{t2}$ | $v_l$ | $v_{t1}$ | $v_{t2}$ |
| ZrTe$_5$ | 6.078 | 3603.05 | 1262.00 | 2391.67 | 2454.35 | 1262.00 | 940.54 | 3531.75 | 2391.67 | 940.45 |
| HfTe$_5$ | 7.039 | 3438.88 | 1106.66 | 2338.24 | 2313.27 | 1106.66 | 757.61 | 3420.42 | 2338.24 | 757.61 |

The obtained value of density for ZrTe$_5$ is consistent with experimental work [20]. To our best knowledge, anisotropic wave velocities are not experimentally measured or theoretically calculated before. This information is important in acoustic device designing.

Using the polycrystalline moduli calculated before, we can calculate longitudinal ($v_l$), transverse ($v_t$) elastic wave velocities and average ($v_m$) sound velocity using the following equations [94]:

$$v_l = \sqrt{\frac{M}{\rho}}, \quad v_t = \sqrt{\frac{G}{\rho}} \quad \text{and} \quad v_m = \left[\frac{1}{3}\left(\frac{2}{v_t^3} + \frac{1}{v_l^3}\right)\right]^{-\frac{1}{3}} \qquad (18)$$

A useful acoustic parameter is radiation factor. It is a useful parameter for soundboard design. For example, in the front plate of violin, we need a material with high radiation factor, whereas in the back plate we need an efficient reflector with comparatively low radiation factor. The ability of a vibrating surface to act as a sound radiator can be measured using this parameter. The intensity, I, of sound generation is proportional to $\sqrt{\frac{G}{\rho^3}}$. This term is known as *radiation factor* [116]. A material's acoustic impedance is a useful parameter in the design of transducers, vehicles, airplane engines, and many other acoustic applications. The amount of acoustic energy that is reflected and transmitted at the interface between two media can be measured by their acoustic impedances. The formula to calculate acoustic impedance is [116]:

$$Z = \sqrt{\rho G} \tag{19}$$

All of these parameters are calculated and shown in Table 9.

**Table 9:** Average longitudinal, transverse, elastic wave velocity ($v_l$, $v_t$, $v_m$ in m/s), radiation factor ($\sqrt{(G/\rho^3)}$ in $m^4$/kg-s) and acoustic impedance (Z in Rayl.) obtained from polycrystalline elastic modulus.

| Material | $v_l$ | $v_t$ | $v_m$ | $\sqrt{(G/\rho^3)}$ | $Z$ (x $10^6$) |
|---|---|---|---|---|---|
| ZrTe$_5$ | 2813.83 | 1648.72 | 1827.97 | 0.27 | 10.02 |
| HfTe$_5$ | 2632.92 | 1512.94 | 1680.34 | 0.21 | 10.65 |

The average sound velocity we calculated, nicely agrees with previous theoretical work [78]. Our calculated values of radiation factor and acoustic impedance suggest that the compounds are efficient reflectors of acoustic energy and can be useful in acoustic devices.

Debye temperature ($\theta_D$) is an important thermophysical parameter because of its connection with thermal conductivity, lattice vibration, interatomic bonding, thermal expansion coefficient, melting temperature etc. The temperature at which all the atomic vibrational modes become active is known as the Debye temperature. $\theta_D$ is related to the characteristic boson energy scale in Cooper pairing associated with the phonons involved in conventional superconducting materials. Larger Debye temperatures indicate lower average atomic mass and stronger interatomic bonding. Large $\theta_D$ also signifies higher melting temperature and higher mechanical wave velocity. It is also related to superconducting transition temperature and highest allowed phonon frequency, $\omega_D$. However, for $T < \theta_D$, quantum nature of vibrational modes are revealed and the higher frequency modes are frozen [117]. We have calculated Debye temperature using this formula [94,116] :

$$\theta_D = \frac{h}{k_B}\left[\left(\frac{3n}{4\pi}\right)\frac{N_A\rho}{m}\right]^{\frac{1}{3}} v_m \tag{20}$$

where $n$ = the number of atoms in the molecule, $\rho$ is the density, $k_B$ =Boltzmann's constant, $M$ = molecular weight, $h$ = Planck's constant, $N_A$ =Avogadro's number and $v_m$ = average sound velocity.

The thermal expansion coefficient ($\alpha_{th}$) is important for epitaxial growth of crystals. It is a useful parameter to understand if the material can be used as thermal barrier coating. Low value of $\alpha_{th}$ of a compound suggests that it is stable over a wide range of temperatures. The formula we have used to calculate thermal expansion coefficient is [118,119]:

$$\alpha_{th} = \frac{\gamma \rho C_v}{3B} \quad (21)$$

with, $C_V \approx C_p$.

Determination of melting temperature is crucial to predict if a material is useful in high temperature applications. From [116] it can be roughly assumed that all classes of crystals experience 2% increase in length by unit increase of temperature. So, we estimated melting temperature of these compounds by the empirical formula given as [116]:

$$\alpha_{th} \approx \frac{0.02}{T_m} \quad (22)$$

The amount of heat absorbed or released for a particular amount of change in temperature of a system is characterized by heat capacity, $C_p$. The larger the heat capacity, the larger is thermal conductivity and the lower is thermal diffusivity. We have calculated volumetric heat capacity here, $\rho C_P$. This parameter calculates how much thermal energy is needed to raise an object's temperature by one degree per unit volume. The formula we have used is [116]:

$$\rho C_p = \frac{3k_B}{\Omega} \quad (23)$$

Here, $\Omega = 1/N$, where, $N$ = atomic number density.

Grüneisen parameter, $\gamma$, is also calculated and tabulated in Table 10. It represents the relative change in the lattice's vibrational frequency brought on by a proportional change in volume. Less anharmonicity and less sensitivity of vibrational modes to volume changes are indicated by lower values of this quantity, whereas higher values suggest that the material's lattice vibrations are strongly impacted by volume changes, frequently producing severe anharmonic effects. It can be calculated using [121]:

$$\gamma = \frac{3(1+v)}{2(2-3v)} \quad (24)$$

Phonons are quanta of lattice vibration. Phonons are linked with a variety of physical and transport properties like heat capacity, thermal conductivity, electrical conductivity etc. The wavelength at which the phonon distribution attains a peak is denoted by $\lambda_{dom}$. The formula to calculate $\lambda_{dom}$ is [122]:

$$\lambda_{dom} = \frac{12.566 v_m}{T} \times 10^{-12} \quad (25)$$

**Table 10:** Debye temperature (θ_D in K), Grüneisen parameter (γ), heat capacity per unit volume ($\rho Cp$ in J/m$^3$-K), thermal expansion coefficient ($\alpha_{th}$ in K$^{-1}$), melting temperature ($T_m$ in K) and the dominant phonon wavelength at 300 K ($\lambda_{dom}$ in m) for ZrTe$_5$ and HfTe$_5$.

| Material | θ_D | $\alpha_{th}$ (x10$^{-5}$) | $T_m$ | $\rho Cp$ (x10$^6$) | γ | $\lambda_{dom}$ (x10$^{-12}$) |
|---|---|---|---|---|---|---|
| ZrTe$_5$ | 169.32 | 2.28 | 876.85 | 1.24 | 1.44 | 76.57 |
| HfTe$_5$ | 157.40 | 2.37 | 841.15 | 1.29 | 1.51 | 70.38 |

Low Debye temperatures suggest soft nature of these compounds, weak chemical bonding, low phonon thermal conductivity.

Low values of heat capacity suggest that these materials' sensitive nature due to temperature change and heat energy exchange [107–109]. The higher values of thermal expansion coefficients and lower values of melting temperatures suggest these compounds are very sensitive to temperature changes, not suitable in very high temperature applications, but can be used in thermal sensors. Values of heat capacities and thermal expansion coefficients are consistent with available experimental results [15,17,19]. The values of Grüneisen parameter suggest intermediate level of anharmonicity [71,73].

Lattice vibration carries heat energy and it is measured by lattice thermal conductivity. It is an important parameter to investigate high temperature applications. Lattice thermal conductivity is calculated using the formula given by Slack [117, 121]:

$$k_{ph} = A(\gamma) \frac{M_{av}\theta_D^3 \delta}{\gamma^2 n^{\frac{2}{3}} T} \tag{26}$$

where, $A(\gamma)$ is a function of Grüneisen parameter [123], $M_{av}$ is average molecular weight, $\theta_D$ is the Debye temperature, n is the number of atoms in conventional unit cell, $\delta$ is the cubic root of atomic volume, $T$ is absolute temperature and $\gamma$ is Grüneisen parameter. The calculated value of $k_{ph}$ at 150 K is tabulated in Table 11. This formula is usually useful when the temperature is less than Debye temperature. Above Debye temperature, phonon-phonon scattering increases, Umklapp scattering dominates which reduces contribution of lattice thermal conductivity in total thermal conductivity.

Thermal conductivity of a compound approaches a minimum ($k_{min}$) at high temperatures well above the Debye temperature. Based on Debye model, we have formulas for thermal conductivity in different directions [123]:

$$k_{min} = \frac{k_B}{2.48} n_v^{\frac{2}{3}} (v_l + v_{t1} + v_{t2}) \tag{27}$$

here, $n_v$ is the number of atoms per unit volume. Isotropic minimum thermal conductivity is calculated using [122]:

$$k_{min} = k_B v_m (V_{atomic})^{\frac{-2}{3}} \tag{28}$$

where $V_{atomic}$ is the cell volume per atom.

**Table 11:** The number of atoms per unit volume ($n$ in m$^{-3}$), isotropic lattice thermal conductivity ($k_{ph}$ in W/m-K) at 150 K, minimum thermal conductivity ($k_{min}$ in W/m-K) in different directions.

| Material | n (x10$^{28}$) | $k_{ph}$ | $k_{min}$ [1 0 0] | $k_{min}$ [0 1 0] | $k_{min}$ [0 0 1] | $k_{min}$ |
|---|---|---|---|---|---|---|
| ZrTe$_5$ | 3.01 | 45.84 | 0.39 | 0.25 | 0.37 | 0.24 |
| HfTe$_5$ | 3.11 | 29.32 | 0.38 | 0.23 | 0.36 | 0.23 |

The tabulated values suggest us that in high temperature, minimum thermal conductivity attained by these compounds are very low. For thermoelectric materials, figure of merit is inversely proportional to thermal conductivity. So, our study suggest that these two compounds may have high efficiency in thermal power generation. They are very good candidates for low temperature thermoelectric applications, as known from previous studies [22,122].

### 3.4. Electronic Properties

### 3.4.1. Electronic Band Structure

We have calculated bulk band structures of these two compounds using both CASTEP and VASP. With VASP, we also calculated band structure incorporating spin-orbit interaction (SOI). Band structure provides a number of information about physical properties of a material. Optical properties of a material are connected with electronic properties revealed by the electronic band structure. At first, we have presented band structures of these compounds calculated using CASTEP. Then we have presented our results calculated via VASP, with and without SOI. The band structures are calculated along high symmetry lines of Brillouin Zone (BZ) suggested by [125] while using CASTEP and along high symmetry lines suggested by [126] while using VASP. All of these results are shown in Figures 4, 5, and 6.

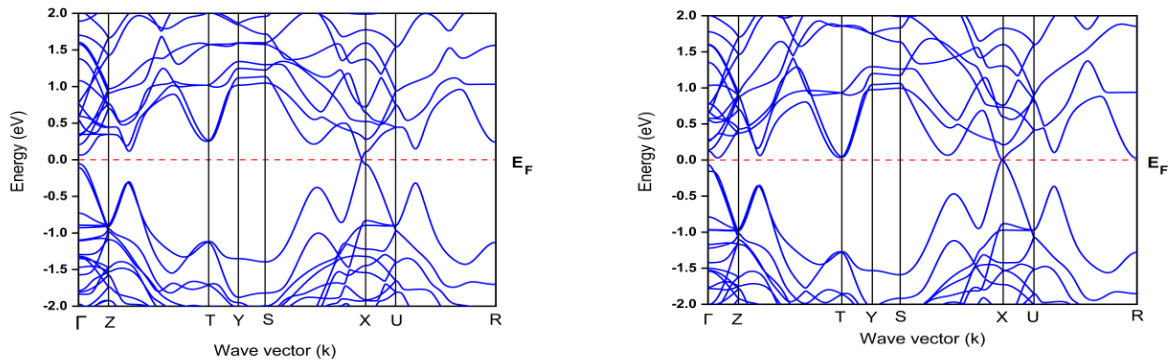

(a)                                                                 (b)

**Figure 4:** Electronic band structures of (a) ZrTe$_5$ and (b) HfTe$_5$ along high symmetry directions using CASTEP for simple orthorhombic structure suggested by [125].

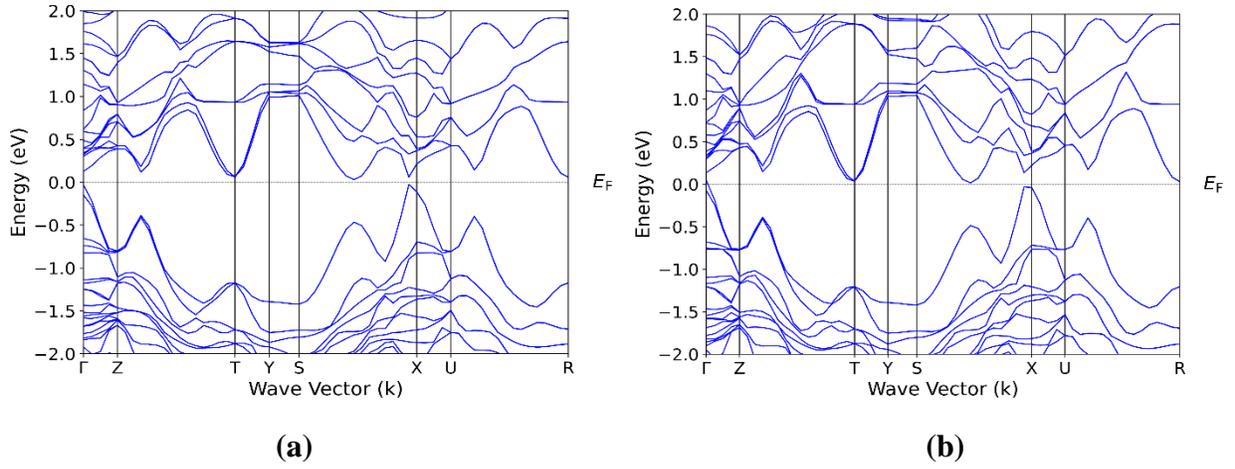

**Figure 5:** Electronic band structures of (a) ZrTe$_5$ and (b) HfTe$_5$ along high symmetry directions using VASP for simple orthorhombic structure suggested by [125].

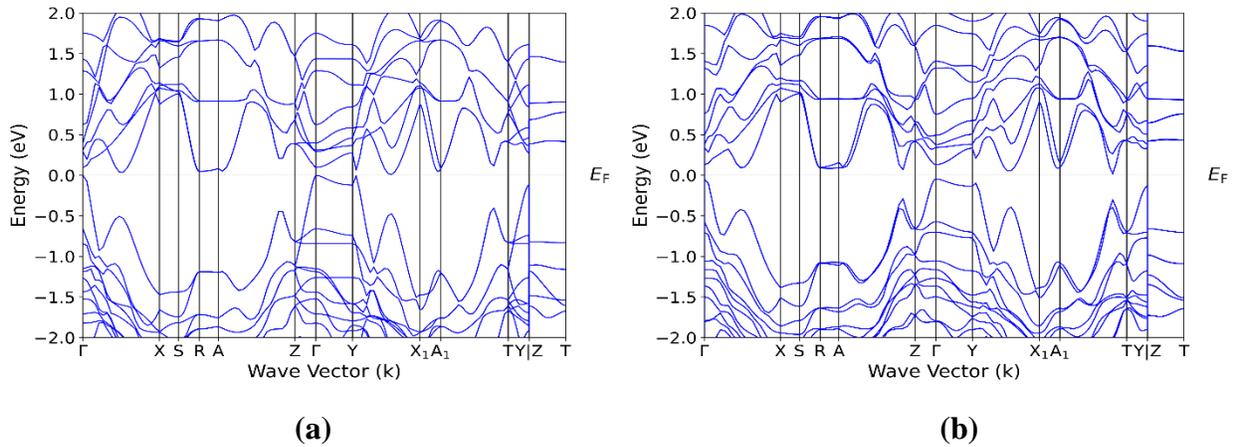

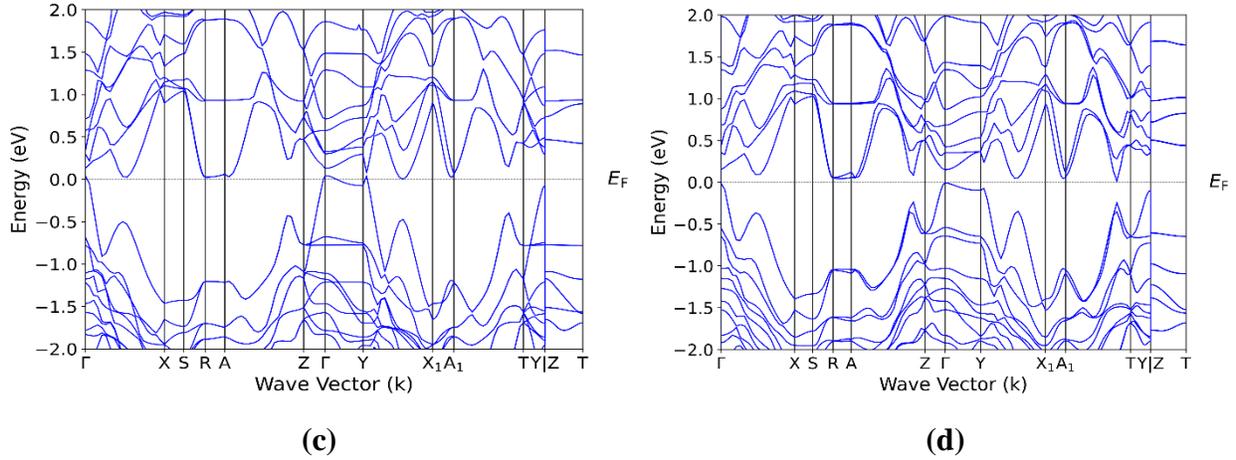

**(c)** **(d)**

**Figure 6:** Electronic band structures of ZrTe$_5$ (a) without SOI (b) with SOI and HfTe$_5$ (c) without SOI (d) with SOI along high symmetry directions using VASP for base centered orthorhombic structure suggested by [126].

In above figures, the Fermi level, E$_F$ is set at 0 eV. Without SOI, CASTEP results suggested semi-metallic nature of these materials (Figure 4). Band touching between valence and conduction states are seen and the dispersion around the respective nodes are almost linear, which bear topological signature of band structures of these compounds. It was not possible to calculate band structures of these compounds with SOI using CASTEP. However, using VASP, we calculated band structures both with SOI and without SOI (Figure 6). Without SOI, for ZrTe$_5$, indirect band gap was 10.7 meV while for HfTe$_5$, we got no indirect band gap. When SOI was incorporated, band gaps we got for ZrTe$_5$ and HfTe$_5$ were 60 meV and 21.6 meV, respectively. The band structures we calculated using VASP for both materials are consistent with previous studies [42,47]. The band gap of HfTe$_5$ is consistent with previous theoretical and experimental studies [47]. Previous experimental studies on ZrTe$_5$ reported band gaps ranging from 6-100 meV [46,48,50,52,125–128]. So, our obtained result is of the correct order and consistent. The resemblance of band structures with previous results [42,47] where Z$_2$ invariant is calculated, also confirm topological nature of these compounds, i.e., they are 3D strong topological insulators (STI). Though, the non-trivial band topology is not due to SOI, instead this is due to non-symmorphic nature of their space group [41]. For these compounds SOI works as a symmetry breaking perturbation, lifts degeneracies and opens up an energy gap around the Fermi level. Flat bands along Y-S direction in CASTEP results (Figure 4) and R-A direction in VASP results (Figure 6) suggest that massive effective charge carriers are present and as a consequence, anisotropy in charge transport will be observed.

### 3.4.2. Density of States

Total electronic density of states (TDOS) and atom resolved partial density of states (PDOS) of both of the compounds are plotted below. We used both CASTEP and VASP for TDOS and PDOS calculations. The Fermi level (E$_F$) is denoted by the vertical line at 0 eV.

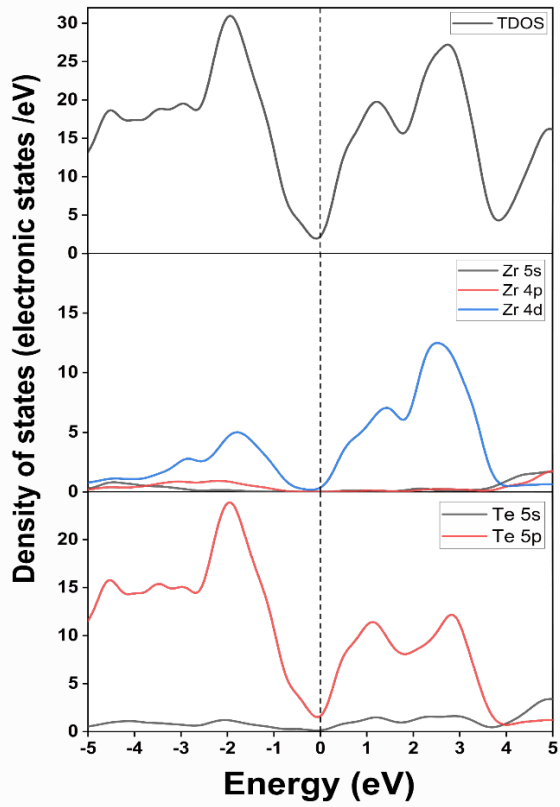 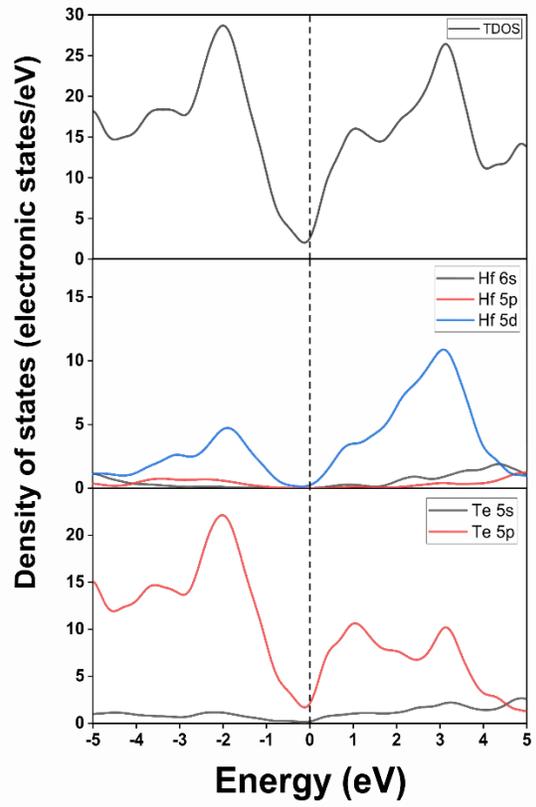

**(a)** **(b)**

**Figure 7:** TDOS and PDOS of (a) ZrTe$_5$ and (b) HfTe$_5$ calculated using CASTEP.

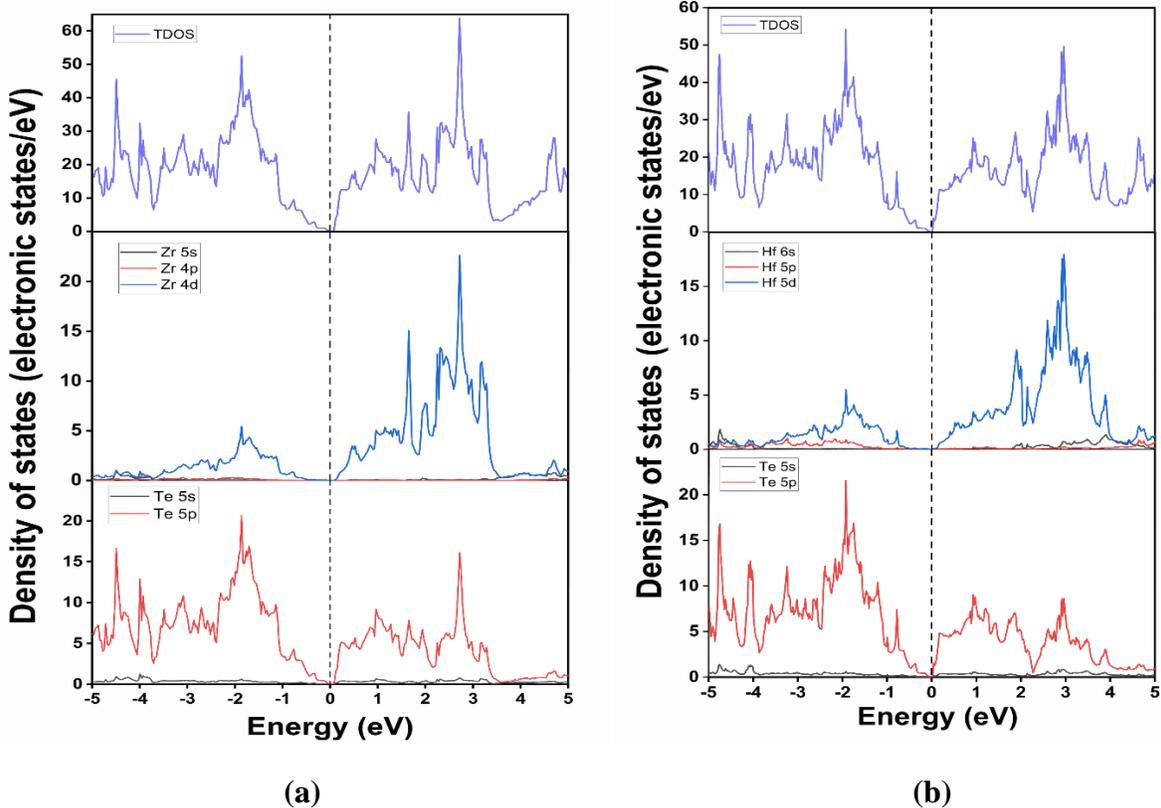

**Figure 8:** TDOS and PDOS of (a) ZrTe$_5$ and (b) HfTe$_5$ calculated using VASP (without SOI).

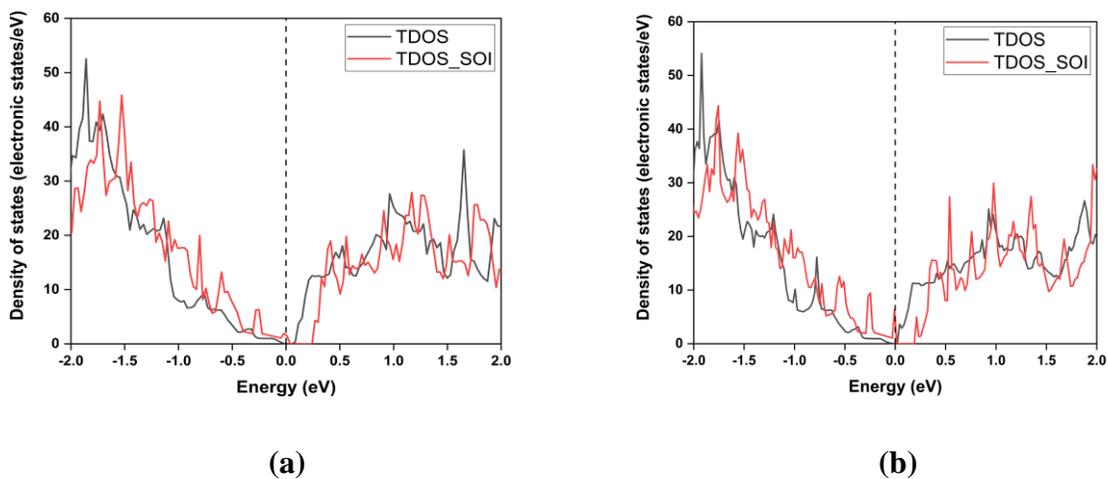

**Figure 9:** TDOS comparison for (a) ZrTe$_5$ and (b) HfTe$_5$ with and without SOI.

CASTEP results are calculated without SOI, they show semi metallic nature, non-zero density of states near Fermi level. The Fermi level is located almost at the center of the pseudogap (Figure 7). This is an indication of high electronic stability [131]. VASP results without SOI show almost similar characteristics. When we compared the results of the total density of states with SOI, we can clearly notice that SOI increase gaps and enhance insulating properties. The gaps are pretty close to Fermi level. Chemical or mechanical processes, for instance, doping or pressure, can be employed to adjust the Fermi level and so we can adjust Fermi level to have zero states. These results reassure us about insulating properties of these compounds. From PDOS plots, it can be confirmed that for both compounds, Te-5p orbital dominates the energy states both below and over Fermi level. The conduction band contribution also prominently come from Zr-4d and Hf-5d orbitals, respectively, for $ZrTe_5$ and $HfTe_5$. So, the conducting carriers mainly consist of Te-5p, Zr-4d and Hf-5d electrons. The covalent bond present in the compounds form due to the hybridization of Te-p and X-d orbitals (X= Zr, Hf). The overall electronic structures of $ZrTe_5$ and $HfTe_5$ are quite similar.

## 3.5. Optical Properties

Predicting a material's response to incident electromagnetic radiation requires an understanding of frequency dependent optical characteristics. The optical characteristics of materials are essential to investigate due to their intimate connections to integrated optics applications like optical modulation, optical communications and information processing, optoelectronics, optical devices etc. Additionally, the frequency-dependent optical characteristics give access to a valuable resource for exploration of the electronic band structure, impurity level states, excitons, specific magnetic excitations, lattice vibrations, and localized defects [131]. In this work, we have calculated several optical properties of $ZrTe_5$ and $HfTe_5$: dielectric function $\varepsilon(\omega)$, refractive index $n(\omega)$, optical conductivity $\sigma(\omega)$, reflectivity $R(\omega)$, absorption coefficient $\alpha(\omega)$, and loss function $L(\omega)$ for electric field polarization in all three directions [1 0 0], [0 1 0] and [0 0 1]. Using CASTEP we calculated complex dielectric function's real part, $\varepsilon_1(\omega)$, and imaginary part, $\varepsilon_2(\omega)$; the real part and the imaginary part of complex refractive index, optical conductivity (real, $\sigma_1(\omega)$ and imaginary, $\sigma_2(\omega)$), reflectivity, $R(\omega)$; the absorption coefficient, $\alpha(\omega)$, and the loss function, $L(\omega)$. Then using VASP, we revisited these properties for both of the compounds. In both cases, we have calculated the properties for incident photon energy up to 25 eV.

The imaginary part of the dielectric constant $\varepsilon_2(\omega)$ is calculated by CASTEP using this formula [132]:

$$\varepsilon_2(\omega) = \frac{2e^2\pi}{\Omega\varepsilon_0} \sum_{k,v,c} \left( |\langle \psi_k^c | \hat{u} \cdot r | \psi_k^v \rangle|^2 \delta(E_k^c - E_k^v - E) \right) \tag{29}$$

Here, $\Omega$ is the volume of the unit cell, $\omega$ is the frequency of the incident photon, $e$ is the electronic charge, $\psi_k^v$ and $\psi_k^c$ are the electronic states of electrons in the conduction and valence bands, respectively, with momentum $\hbar k$, **u** is the direction of incident electric field polarization. VASP also calculates this expression using Green-Kubo formula [133]:

$$\varepsilon_{\alpha\beta}^{(2)}(\omega) = \frac{4\pi^2 e^2}{\Omega} \lim_{q \to 0} \left(\frac{1}{q^2}\right) \sum_{c,v,k} \left( 2\omega_k \delta(E_{ck} - E_{vk} - \omega) \times \langle u_{ck+e_\alpha q} | u_{vk} \rangle \langle u_{vk} | u_{ck+e_\beta q} \rangle \right) \tag{30}$$

where, $\varepsilon_{\alpha\beta}$ is defined as: $D_\alpha(\omega) = \varepsilon_{\alpha\beta}(\omega) E_\beta(\omega)$.

$D_\alpha(\omega)$ is displacement vector's αth component. $|u_\alpha\rangle$ corresponds to Bloch state with momentum $\hbar\alpha$, while the real part of $\varepsilon$ is calculated using Kramers-Kronig relations [134]. Once we have real and imaginary part of dielectric functions, we can calculate all frequency dependent optical properties [128,130,133]. For CASTEP calculations, we haven't incorporated any Drude correction, treating the system as non-metallic. For both CASTEP and VASP, we didn't include SOI as the energy gap it opens, are of meV order. The plots are presented below:

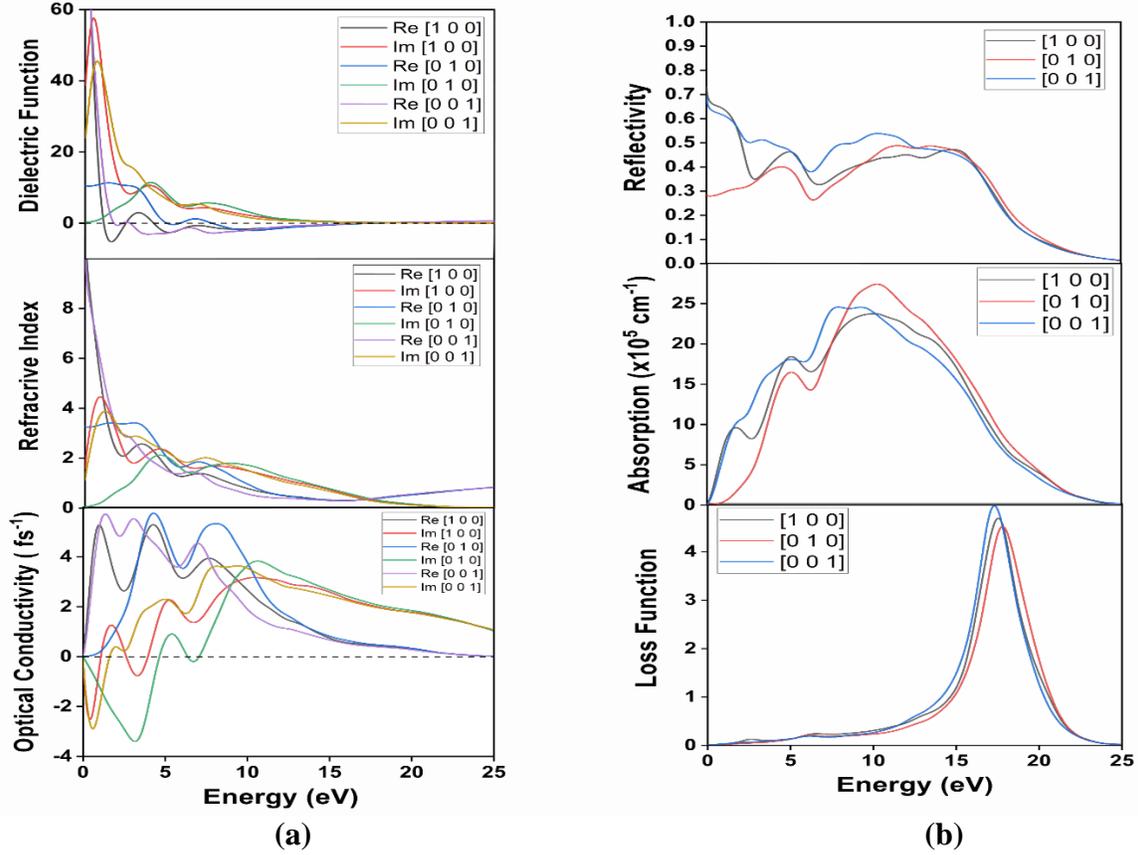

.                              **(a)**                             **(b)**

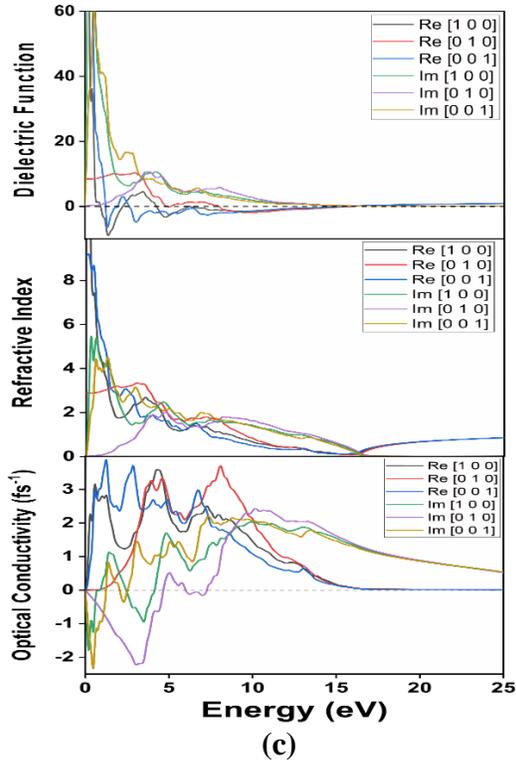
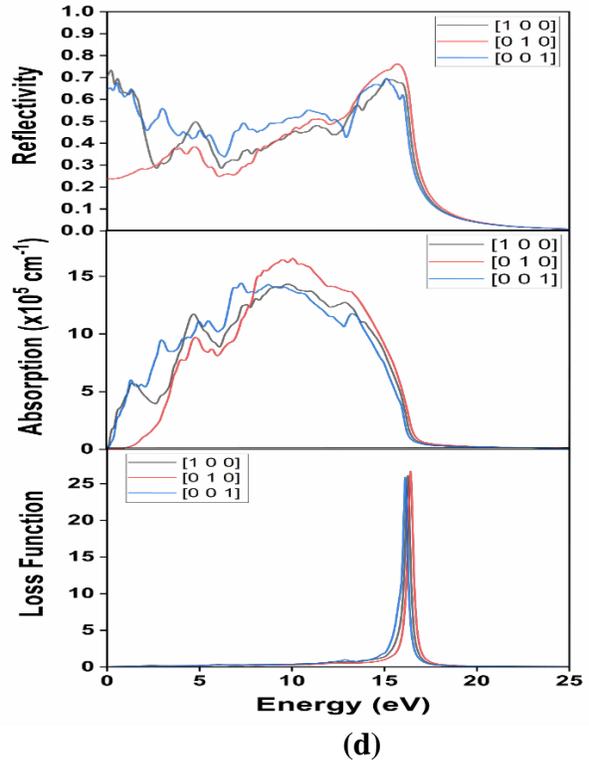

(c)

(d)

**Figure 10:** Optical properties of ZrTe$_5$ calculated using (a)-(b): CASTEP and (c)-(d):VASP.

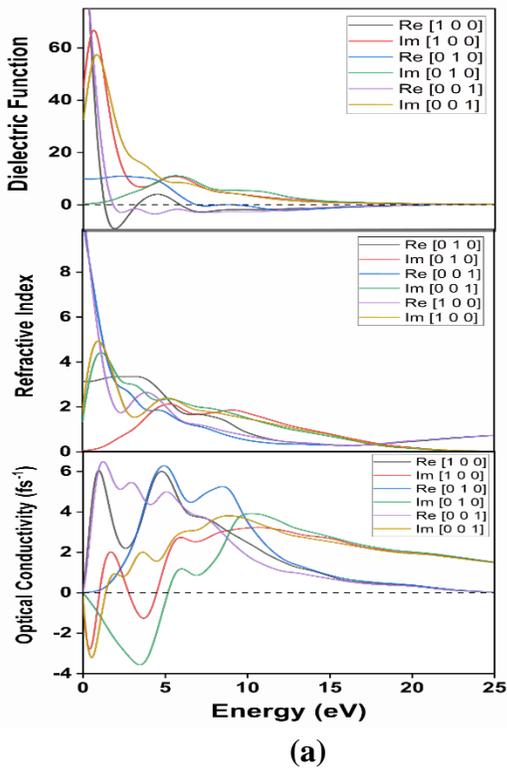
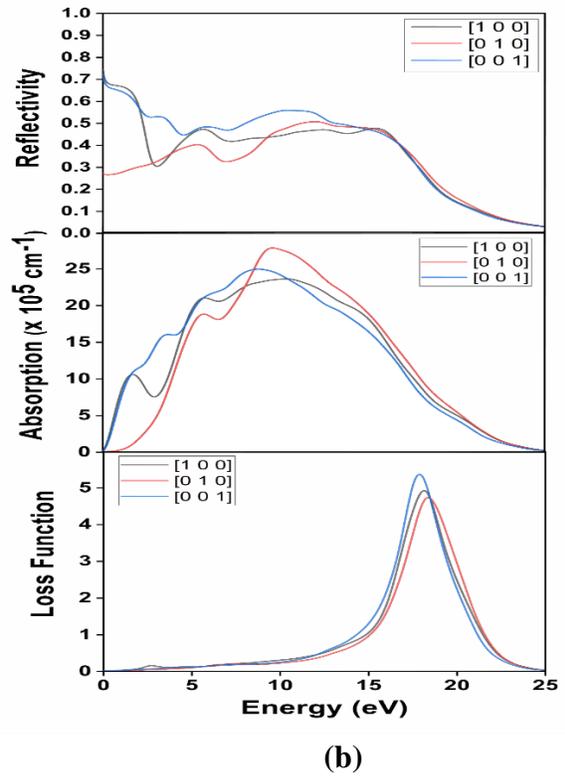

. (a)

(b)

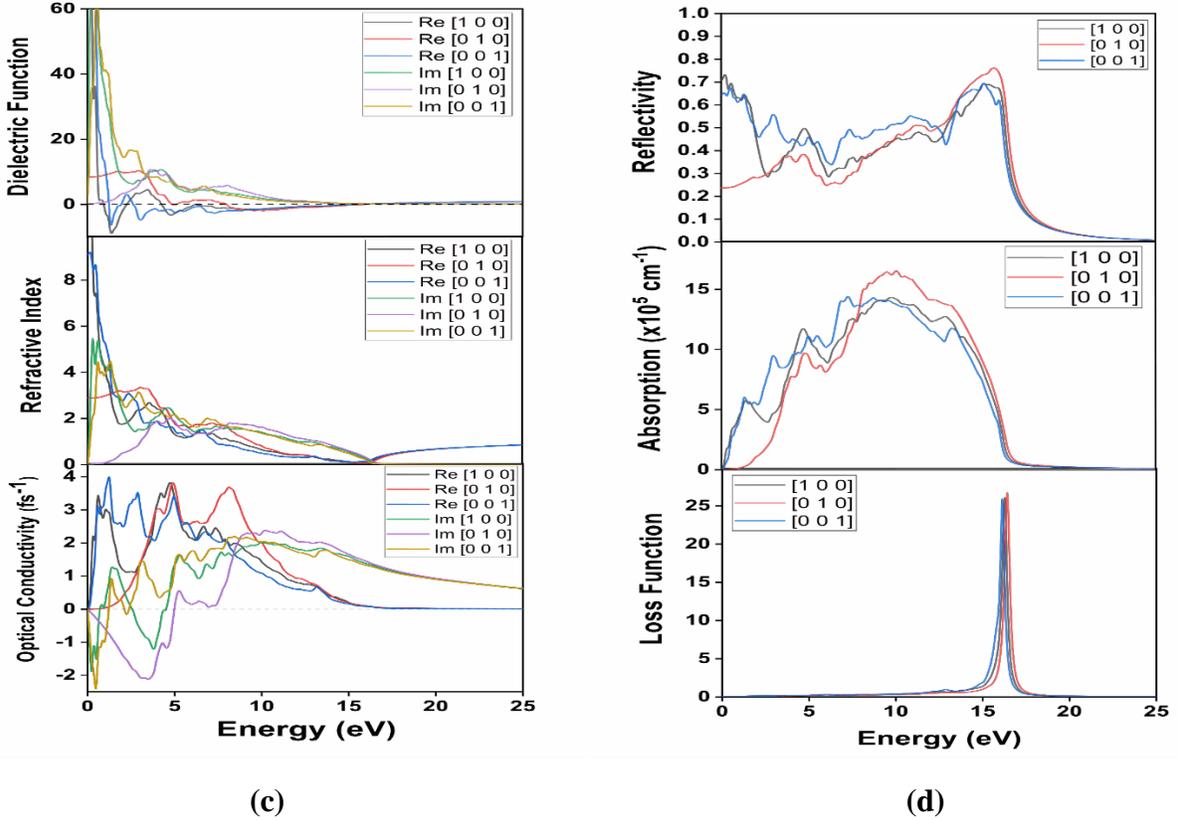

**Figure 11:** Optical properties of HfTe$_5$ calculated using (a)-(b): CASTEP and (c)-(d): VASP.

The real part of dielectric function is connected directly to the electrical polarization of materials, whereas the imaginary part is associated with dielectric loss. $\varepsilon_2(\omega)$ is also linked with electronic band structure and DOS profile. Contribution in dielectric function comes both from interband and intraband optical transitions [136]. Refractive index $n(\omega)$ is associated with phase velocity of electromagnetic (em) wave through the compound. Extinction coefficient, $k(\omega)$ is associated with attenuation of em wave. Optical conductivity $\sigma(\omega)$ provides information about the conduction of free charge carriers over a range of photon frequencies and dynamic response of mobile charge carriers.

Our study suggests that both of these compounds are highly optically anisotropic. Specially, when electric field polarization is along crystallographic **b** direction, all of the properties show significant anisotropy compared to other two directions. The real part of dielectric function shows distinctive phenomena. For all three directions, there is a particular range where $\varepsilon_1(\omega)$ is negative. This range is the forbidden propagation range or "Restrahlen band" for each three directions [137]. In the infrared and visible region, phonon-polaritons have significant contribution in optical response of compounds [138]. The upper and lower bound of the forbidden band is usually decided by the frequency of longitudinal and transverse optical phonons. Optical response in mid UV region is decided by both phonon-polaritons and plasmon polaritons [138,139]. The peaks of $\varepsilon_2(\omega)$ approximately near 1 eV and 3 eV are associated with the DOS peaks seen in Figure 7 below and above the Fermi level.

Very high refractive index in low energy region which corresponds to visible spectrum, suggest that these compounds are good candidate for optoelectronic display devices. Enhanced optical conductivity at low photon energies is a consequence of photo generated electron and holes.

Reflectivity spectra show that these compounds are good reflectors of photon in the visible region for the [100] and [001] electric field polarizations. The reflectivity in the infrared and visible region is significantly lower for the [010] polarization. This anisotropy in the reflectivity is large and can be used to detect polarization states of incident light. Absorption spectra show that these compounds are good absorbers of photon energy specially in the ultraviolet region. These results suggest that these compounds can be useful in solar energy coating devices and bacteria disinfection applications.

Loss spectrum peak corresponds to characteristic plasmon resonance. In non-metallic system, plasmon resonance corresponds to natural frequency of collective oscillation of bound electrons. Plasmon-polaritons have dominant influence on optical response of semiconductors and insulators usually in the ultraviolet region. CASTEP results show a peak near 17-18 eV. But the peak isn't sharp. In CASTEP, we need to specify plasma frequency to incorporate Drude correction. It is important in metallic system. So, we haven't incorporated any Drude correction here. This may be a reason of broadening of width of the peaks in CASTEP results. VASP doesn't require any semi-empirical parameter to calculate optical properties. VASP results show a sharp peak near 16 eV in all three directions for both the compounds. It is consistent with reflectivity and absorption spectrum, both showing sharp fall at this energy. Extinction coefficients are also zero above this particular energy. So, for photon energy greater than this value, the materials will be optically transparent.

## 4. Conclusion

In this work we have investigated a number of physical properties of topological insulators, $ZrTe_5$ and $HfTe_5$. Our study explored a lot of previously uncharted properties of these materials. Their largely anisotropic nature of elastic, mechanical and optical properties are revealed. Moreover, their soft, brittle nature and possible application in mechanical and acoustic devices are investigated. Both these compounds are highly machinable. Several other thermophysical properties have been studied for the very first time. Low Debye temperature and melting temperature suggest their unsuitability in very high temperature applications. Thermal expansion coefficients, heat capacities we calculated are consistent with experimental studies. The minimum thermal conductivities of these materials are very low, indicating that they have potential to be used as thermal barrier coating at low temperatures. Significant reflectivity and absorption over a broad range of energy suggested their application in optoelectronic devices and solar reflectors. Plasmon resonance frequency and forbidden propagation range for propagation in different directions have been calculated. Though, there were a lot of studies on the electronic and topological properties of these compounds, disparities in different studies have always opened up scope and engrossment over these properties. Our calculated band structures and band gaps are consistent with earlier investigations and including SOI, we were able to justify their insulating nature. Another fascinating phenomenon about these two compounds is that their elastic, electronic and optical behavior are almost exactly similar in spite of consisting two different elements. This is due to the nearly similar orbital characteristics of Zr $4d^2 5s^2$ and Hf $5d^2 6s^2$ electrons. Due to interesting electronic and optical properties, a large number of researches on application of topological insulators are going on. We believe that our study will inspire material scientists for further theoretical and experimental investigations on these compounds.

**Data availability**

The data sets generated and/or analyzed in this study are available from the corresponding author on reasonable request.

**Competing Interests**

The authors declare no competing interests.

2024).